\documentclass[pra,preprint,showpacs]{revtex4}
\usepackage[centertags]{amsmath}
\usepackage{amsfonts}
\usepackage{amssymb}
\usepackage{amsthm} 
\usepackage{newlfont}
\usepackage{epsfig}
\usepackage{amscd}
\usepackage{graphicx}
\usepackage{epsfig}
\usepackage{footnote}
\usepackage{color}
\usepackage{xcolor}
\usepackage{graphicx}
\usepackage{subfig}
\usepackage{amscd}
\newcommand{\NN}{{\mathbb N}}

\newcommand{\beq}{\begin{equation}}
\newcommand{\eeq}{\end{equation}}
\newcommand{\ba}{\begin{array}}
\newcommand{\ea}{\end{array}}
\newcommand{\bea}{\begin{eqnarray}}
\newcommand{\eea}{\end{eqnarray}}

\begin{document}

\begin{center}
{\large \sc \bf {Coherence evolution and transfer supplemented by the state-restoring}
}

\vskip 15pt

{\large 
E.B.Fel'dman and A.I.~Zenchuk 
}

\vskip 8pt

{\it $^2$Institute of Problems of Chemical Physics, RAS,
Chernogolovka, Moscow reg., 142432, Russia}.

\end{center}


\begin{abstract}

The evolution of quantum coherences comes with a set of conservation laws provided that the Hamiltonian governing this evolution conserves the spin-excitation number. At that, coherences do not intertwist during the evolution. Using the transmission line and the receiver in the initial ground  state  we can transfer the coherences  to the receiver without interaction between  them, {  although the matrix  elements contributing to each particular coherence intertwist in the receiver's state. }
Therefore we propose a tool based on the unitary transformation at the receiver side to { untwist these elements and thus}  restore    (at least partially)  the structure of the sender's initial density matrix. A communication line with two-qubit sender and receiver is considered as an example of implementation of this technique.

\end{abstract}

\maketitle

\section{Introduction}
\label{Section:Introduction}

The multiple quantum (MQ) NMR dynamics is a basic tool of well developed MQ NMR spectroscopy   studying the nuclear spin distribution 
in different systems \cite{BMGP,DMF}. 
{ Working with spin polarization we essentially deal with the diagonal elements of the density matrix. However, the MQ NMR method allows us to split the whole density matrix into $N+1$ parts, and each of these parts contributes into a specific observable quantity called coherence intensity.}  
Thus studying the coherence intensities and the methods of manipulating  them becomes an important direction in development of  MQ NMR methods. For instance, the problem of relaxation of  MQ coherences was studied in \cite{KS1,KS2,AS,CCGR,BFVV}. A similar problem in nonopore was considered in \cite{DFZ}). 

In MQ NMR experiment, the special sequence of the magnetic pulses is used to generate the so-called two-spin/two-quantum Hamiltonian  ($H_{MQ}$)
which is the non-secular part of  the dipole-dipole interaction  Hamiltonian averaged over fast oscillations.  
It was shown in the approximation of  nearest-neighbor interactions  that  the $H_{MQ}$ Hamiltonian can be reduced to  the  flip-flop XX-Hamiltonian ($H_{XX}$)  \cite{Mattis} via the unitary transformation \cite{DMF}. Notice, that $H_{MQ}$ does not commute with 
the  $z$-projection of the total 
spin momentum $I_z$, while $[H_{XX},I_z]=0$.  

In this paper we consider the evolution  problem for the created MQ coherences.
 Therefore, after creating the coherences, we switch off the irradiation and allow the coherences  to evolve independently  under the Hamiltonian commuting with $I_z$ (this can be, for instance, $H_{dz}$  Hamiltonian \cite{Abragam,Goldman} or $H_{XX}$ flip-flop Hamiltonian).   
 We show that the  coherences do not interact during the evolution governed by the Hamiltonian conserving the $z$-projection of the total spin momentum. This fact gives rise to the set of conservation laws associated with such dynamics, namely, the 
 coherence intensity of an arbitrary order conserves.
But the density-matrix elements contributing into the same order coherence do intertwist. 

In addition, the coherences, created in some subsystem (sender)  can be transferred to another subsystem (receiver) through the transmission line 
without interaction between coherences if only the both receiver and  transmission line are in the initial state having only 
the zero-order coherence.
This process can be considered as a particular implementation of the remote state creation in spin systems \cite{Z_2014,BZ_2015}.
We show that the sender's density-matrix elements in the receiver's state can be  untwisted using the  method 
based on the unitary transformation of the receiver or, more effectively, of the extended receiver.  The theoretical arguments are supplemented with the particular model of communication line having  two-node sender and receiver. Notice that the extended receiver was already used in the  previous papers concerning the remote state creation \cite{BZ_2016} with the purpose of  proper correcting the created state of the receiver and improving the characteristics of the remote state creation \cite{Z_2014,BZ_2015}.

The paper is organized as follows. In Sec.\ref{Section:DC} we select  the matrices $\rho^{(n)}$ responsible for forming 
the $n$-order coherence intensity and study some  extremal values of coherence intensities.
The evolution of the coherence intensities is considered in Sec.\ref{Section:ev}. The transfer of the coherences from the sender to the receiver is studied in Sec.\ref{Section:cohtr}. 
In Sec.\ref{Section:model} we apply the results of previous sections to   a particular model of a chain with 2-qubit sender and receiver.
The brief discussion of obtained results is given in Sec.\ref{Section:conclusion}.

\section{Density matrix and  coherences}
\label{Section:DC}
It was shown { (for instance, see \cite{FL})} that the density matrix of a quantum state can be written as a sum 
\begin{eqnarray}\label{RhoC}
\rho = \sum_{n={-N}}^N  \rho^{(n)},
\end{eqnarray}
where each submatrix $ \rho^{(n)}$  consists of the elements of $\rho$ responsible for the spin-state  transitions  changing the total $z$-projection of the spin momentum by $n$. These elements contribute to the so-called $n$-order coherence intensity $I_n$ which can  be registered using the MQ NMR methods.  To select the density matrix elements contributing to the $n$-order coherence we turn to the { density-matrix representation in the multiplicative basis
\begin{eqnarray}\label{multb}
|i_1\dots i_N\rangle,\;\;i_k=0,1,\;\;k=1,\dots,N,
\end{eqnarray}
where $i_k$ denotes the state of the $k$th spin.
Thus, the transformation from the computational basis to the multiplicative one reads}
\begin{eqnarray}\label{mult}
\rho_{ij}= \rho_{i_1\dots i_N;j_1\dots j_N},\;\;\; i=\sum_{n=1}^N i_n 2^{n-1} +1,\;\; j=\sum_{n=1}^N j_n 2^{n-1} +1.
\end{eqnarray}
Then, 
according to  the definition, 
\begin{eqnarray}\label{defI}
I_n(\rho) ={\mbox{Tr}} \Big(\rho^{(n)}\rho^{(-n)}\Big) =  \sum_{\sum_k (j_k - i_k) = n} |\rho_{i_1\dots i_N;j_1\dots j_N}|^2,\;\;
|n|\le N.
\end{eqnarray}

\subsection{Extremal values of  coherence intensities}

First of all we find the extremal values of the zero order coherence intensity of $\rho$  provided that all other 
coherences absent, so that $\rho=\rho_0$. By the definition (\ref{defI}),
\begin{eqnarray}
I_0={\mbox{Tr}} \Big(\rho_0 \rho_0\Big) = {\mbox{Tr}} \left(U_0\Lambda_0 U_0^+\right)^2 = {\mbox{Tr}} \Lambda_0^2 =
\sum_{i=1}^{2^N} \lambda_{0i}^2,
\end{eqnarray}
where $N$ is the number of spins in the sender,  $\Lambda_0={\mbox{diag}}(\lambda_{01},\dots,\lambda_{02^N})$ and $U_0$ are, respectively, the 
eigenvalue and eigenvector matrices of $\rho$. 
Therefore we have to find the extremum of $I_0$ with the normalization condition
$\sum_{i=1}^{2^N} \lambda_{0i} =1$. 
Introducing the Lagrange factor $\alpha$ we 
reduce the problem to constructing the extremum of the function
\begin{eqnarray}
\tilde I_0 = \sum_{i=1}^{2^N} \lambda_{0i}^2 - \alpha \left( \sum_{i=1}^{2^N} \lambda_{0i} -1\right).
\end{eqnarray}
Differentiating with respect to $\lambda_{0i}$ and equating the result to zero we obtain the system of equations
\begin{eqnarray}
2\lambda_{0i}=\alpha,\;\;i=1,\dots,2^N,
\end{eqnarray}
therefore, $\lambda_{0i}=\frac{\alpha}{2}$. Using the normalization we have $\alpha=\frac{1}{2^{N-1}}$, 
so that $\lambda_{0i}=\frac{1}{2^N}$. The second derivative of $\tilde I_0$ shows that this is a minimum. Thus, we have
\begin{eqnarray}
I_0^{min}=\frac{1}{2^N}, \;\;\rho|_{I_{0}^{min}} = \frac{1}{2^N}E,
\end{eqnarray}
where $E$ is the $2^N\times 2^N$ identity matrix.
To find the maximum value of $I_0$ we observe that 
\begin{eqnarray}
\sum_{i=1}^{2^N} \lambda_{0i}^2 =\left(\sum_{i=1}^{2^N} \lambda_{0i}\right)^2 -\sum_{i\neq j} \lambda_{0i}\lambda_{0j}=1-\sum_{i\neq j} \lambda_{0i}\lambda_{0j} \le 1. 
\end{eqnarray}
It is obvious that the unit can be achieved if there is only one nonzero eigenvalue $\lambda_{01}=1$.
Thus 
\begin{eqnarray}
I_0^{max}=1, \;\;\rho|_{I_{0}^{max}} = {\mbox{diag}}(1,\underbrace{0,0,\dots0}_{2^N-1}).
\end{eqnarray}

Now we proceed to the analysis of the $n$-order coherence intensity for the matrix having only three non-zero  coherences of  zero- and  $\pm n$-order, 
assuming that  the zero-order coherence intensity $I_{0}$ is minimal, i.e., 
\begin{eqnarray}\label{rhoin}
\rho=\frac{1}{2^N}E + \tilde \rho^{(n)}  = U_n \left(\frac{1}{2^N}E +\Lambda_n\right) U^+_n,\;\;\;\tilde \rho^{(n)}=\rho^{(n)} + \rho^{(-n)}
\end{eqnarray}
where $\Lambda_n={\mbox{diag}}(\lambda_{n1},\dots,\lambda_{n2^N})$ and $U_n$ are the matrices of eigenvalues and eigenvectors of 
$\tilde \rho^{(n)}$. Of course, $U_n$ is also the  eigenvector matrix for the whole $\rho$ in this case and
\begin{eqnarray}\label{constr2}
\sum_{i=1}^{2^N} \lambda_{ni} =0.
\end{eqnarray}

Now  we proof one of the interesting property of the eigenvalues for the considered case.

{\bf  Proposition 1.}
Eigenvalues  $\lambda_{ni}$ appear in pairs:
\begin{eqnarray}\label{pairs}
\lambda_{n(2i-1)}= \eta_{ni}, \;\;\lambda_{n(2i)}= -\eta_{ni},
\;\;\;i=1,\dots,2^{N-1}. 
\end{eqnarray}

{\it Proof.} First we show that,
along with $\tilde \rho^{(n)}$, the odd powers of  this matrix  are also traceless. 
For instance, let us show that 
\begin{eqnarray}\label{rr}
{\mbox{Tr}}(\tilde \rho^{(n)})^3 = \sum_{i,j,k} \tilde \rho^{(n)}_{ij} \tilde \rho^{(n)}_{jk} \tilde \rho^{(n)}_{ki} = 0.
\end{eqnarray}
 Using the multiplicative basis for the density-matrix elements in the rhs of eq. (\ref{rr}),  we remark that only such elements $\tilde \rho_{ij}$, $\tilde \rho_{jk}$  and 
$\tilde \rho_{ki}$ are nonzero that, respectively, 
$\sum_m i_{m} -\sum_m j_{m} = \pm n$, $\sum_m j_{m} -\sum_m k_{m} = \pm n$ and $\sum_m k_{m} -\sum_m i_{m} = \pm n$. However,  summing all these equalities 
we obtain the identical  zero in the lhs and either $\pm 3 n$ or $\pm n$ in the RHS. 
This contradiction means that there must be zero matrix elements in each term of the sum (\ref{rr}), i.e., the trace is zero. 

Similar consideration works for higher odd powers of $\tilde \rho^{(n)}$
(however, the sum $\tilde \rho^{(n)} + \tilde \rho^{(k)}$, $k\neq n$, doesn't possesses this property, i.e., the trace of any its power is non-zero in general).
Consequently, along with (\ref{constr2}), the following equalities hold:
\begin{eqnarray}\label{sumni}
\sum_{i=1}^{2^N} \lambda_{ni}^m =0 \;\;{\mbox{for any odd}}\;\;m.
\end{eqnarray}
Condition (\ref{sumni}) holds for any odd $m$ if only  the eigenvalues $\lambda_{ni}$ appear in pairs (\ref{pairs}). 
{To prove this  statement, first we assume that all eigenvalues are non-degenerate and 
let the eigenvalue $\lambda_{n1}$ be maximal by absolute value.
We divide  sum (\ref{sumni}) by $\lambda_{n1}^m$:
\begin{eqnarray}\label{sumni2}
1+\sum_{i=2}^{2^N} \left(\frac{\lambda_{ni}}{\lambda_{n1}}\right)^m =0, \;\;{\mbox{for odd}}\;\;m.
\end{eqnarray}
Each term in the sum can not exceed one by absolute value. Now we take the limit 
$m\to\infty$ in eq.(\ref{sumni2}). It is clear that   all the terms such that $\left|\frac{\lambda_{ni}}{\lambda_{n1}}\right|<1$ 
 vanish. Since this sum is zero, there must be an eigenvalue $\lambda_{n2}$ such that  $\lambda_{n2} = -\lambda_{n1}$. Then, the 
appropriate term in (\ref{sumni2}) yields -1. So, two first terms in sum (\ref{sumni2}) cancel each other which reduces 
(\ref{sumni2}) to
\begin{eqnarray}\label{sumni3}
\sum_{i=3}^{2^N} \lambda_{ni}^m =0, \;\;{\mbox{for odd}}\;\;m.
\end{eqnarray} 
Next, we select the maximal (by absolute value) of the remaining eigenvalues, repeat our arguments and conclude that there are two more eigenvalues equal by absolute value and having opposite signs. And so on. Finally, after $2^{N-1}$,  steps we result in  conclusion that all eigenvalues appear in pairs (\ref{pairs}).

Let  the $(2k+1)$th eigenvalue on the $(2k+1)$-step is $s$-multiple, i.e. $\lambda_{n(2k+1)} =\dots = \lambda_{n(2k+s)}$. Then the sum (\ref{sumni}) gets the form 
\begin{eqnarray}\label{sumni4}
\sum_{i=2k+1}^{2^N} \left(\frac{\lambda_{ni}}{\lambda_{n(2k+1)}}\right)^m = 
s +\sum_{i=2k+s+1}^{2^N} \left(\frac{\lambda_{ni}}{\lambda_{n(2k+1)}}\right)^m,\;\; s\in\NN,\;\;s\le N-2k,\;\;{\mbox{odd}} \;\;m.
\end{eqnarray}
Now, to compensate $s$ we need   an $s$-multiple   eigenvalue, such that   
$\lambda_{n(2k+s+1)} = \dots = \lambda_{n(2k+2s)} = - \lambda_{n(2k+1)}$. Thus, if  there is $s$-multiple positive eigenvalue, there must be an $s$-multiple negative eigenvalue. 
This ends the proof.} $\Box$

Next, since all the eigenvalues of $\rho$ must be non-negative and the density matrix $\rho$ has the structure (\ref{rhoin}), the negative eigenvalues $\eta_{ni}$ can not exceed $\frac{1}{2^N}$ by absolute value. Therefore, the  maximal $n$-order coherence intensity  corresponds to the case 
\begin{eqnarray}
\eta_{ni} =\frac{1}{2^N}.
\end{eqnarray}
Consequently, 
\begin{eqnarray}
I_n^{max}+I_{-n}^{max} = 2 I_n^{max} =\sum_{j=1}^{N_n} \lambda_{ni}^2 =\frac{N_n}{2^{2N}}\le \frac{1}{2^N}, 
\end{eqnarray}
where $I_n^{max}=I_{-n}^{max}$  and  $N_n$ is the number of nonzero eigenvalues of $\tilde \rho^{(n)}$. 
This number equals to the rank of $\tilde \rho^{(n)}$ which, in turn,  can be found as follows.  

{\bf  Proposition 2.}  The rank of the matrix $\tilde \rho^{(n)}$ can be calculated using the formula
\begin{eqnarray}\label{ran}
N_n={\mbox{ran}}\;\tilde \rho^{(n)} = \sum_{k=0}^{N} \min \left(
\left(N\atop k \right) ,\left(N\atop k+n \right)+\left(N\atop k-n \right) \;\; 
\right),
\end{eqnarray}
where 
the binomial coefficients  $\left(N\atop m \right)=0$ for $m<0$.

{\it Proof.}
For the $n$-order coherence, the number of states with $k$ excited spins equals 
$ \left(N\atop k \right)$. The $\pm n$-order coherence collects the elements of $\rho$ responsible for  transitions from the states with  
$k$ excited spins to the states with 
$k\pm n$ excited spins. All together, there are $\left(N\atop k+n \right)+\left(N\atop k-n \right)$ such transitions. 
These transitions can be collected into the matrix  of $ \left(N\atop k \right)$ columns and 
$\left(N\atop k+n \right)+\left(N\atop k-n \right)$ rows, whose maximal rank equals $\min \left(
\left(N\atop k \right) ,\left(N\atop k+n \right)+\left(N\atop k-n \right)\right)$. 
Obviously, the rank of $\tilde \rho^{(n)}$ equals the sum of calculated ranks for different $k=0,\dots,N$, i.e., we obtain formula (\ref{ran}).$\Box$

{\bf Consequence.}
For the coherence intensity of the first order ($n=1$)   eq.(\ref{ran}) yields:
\begin{eqnarray}\label{ran1}
N_1= \sum_{k=0}^{N} 
\left(N\atop k \right) = 2^N.
\end{eqnarray}

{Proof.}
We have to show that in this case 
\begin{eqnarray}\label{con}
\left(N\atop k \right) \le \left(N\atop k+1 \right)+\left(N\atop k-1 \right),\;\;0\le k \le N.
\end{eqnarray}
First we consider the case $k>1$ and $k<N$. Then 
\begin{eqnarray}\label{intermed1}
\left(N\atop k+1 \right)+\left(N\atop k-1 \right) = \left(N\atop k \right) \left(\frac{N-k}{k+1} + \frac{k}{N-k+1}\right).
\end{eqnarray}
Let us show that the expression inside the parenthesis is $\ge 1$.
After simple transformations, this condition takes the form 
\begin{eqnarray}\label{ge}
3 k^2 - 3 N k +N^2 -1\ge 0,
\end{eqnarray}
where the lhs is a quadratic expression in $k$. The roots of the lhs read
\begin{eqnarray}
k_{1,2}=\frac{3 N \pm\sqrt{12-3 N^2}}{6}, 
\end{eqnarray}
which are imaginary for $N>2$.  Therefore the parabola $3 k^2 - 3 N k +N^2$ lies in the upper half-plane $k$
for $N> 2$ and consequently condition (\ref{ge}) holds for $N\ge2$. In our case, the minimal $N$ is 2, which corresponds to the 1-qubit sender and 1-qubit receiver without the transmission line between them.

If $k=1$ then, instead of (\ref{intermed1}), we have 
\begin{eqnarray}
\left(N\atop 2 \right)+\left(N\atop 0 \right) = \left(N\atop 2 \right) +1 = \left(N\atop 1 \right) \frac{N-1}{2} +1 \ge \left(N\atop 1 \right),\;\;N\in\NN .
\end{eqnarray}
 Therefore condition (\ref{con}) is  also satisfied.

If $k=0$, then $\left(N\atop 1 \right)=1$ and 
\begin{eqnarray}
\left(N\atop 1 \right)+\left(N\atop -1 \right) = \left(N\atop 1 \right) >\left(N\atop 0 \right) ,
\end{eqnarray}
therefore condition (\ref{con}) is also satisfied. 

The cases $k=N$ can be considered in a similar way.
$\Box$

Thus, $N_1$ equals 
the maximal possible  rank $N_1={\mbox{ran}} \;\tilde \rho^{(1)}$, so that $\displaystyle 2 I_1^{max}= \frac{1}{2^{N}}$.
{ Similarly, for the $N$-order coherence we have only two nonzero terms in (\ref{rr}) which give  $N_N=2$ and 
$2 I_N^{max} =\frac{1}{2^{2N-1}}$.
For the  intensities of the other-order coherences we do not give similar result for any $N$. The 
maximal coherence intensities of the $n$-order ($n>0$) for $N=2,\dots,5$ are given in Table \ref{Table1}.}
This table shows the ordering of $I_n^{max}$: 
\begin{eqnarray}\label{order}
I_0^{max} > I_1^{max}> \dots >I_N^{max}.
\end{eqnarray}
\begin{table}
\begin{tabular}{|c|cc|ccc|cccc|ccccc|}
\hline
$N$ & \multicolumn{2}{|c|}{2} & \multicolumn{3}{|c|}{3}&\multicolumn{4}{|c|}{4}&\multicolumn{5}{|c|}{5}\cr
\hline
$n$ &              1             & 2        &   1             & 2 &3    &              1            & 2 &3 &4 &1&2&3&4&5 \cr
$N_n$&             4             & 2        &   8            & 4 &2&             16            & 12 &4&2&32&24&14&4&2\cr
$2 I_n^{max}$&$\displaystyle \frac{1}{4}$    & $\displaystyle \frac{1}{8}$&   $\displaystyle \frac{1}{8}$    & $\displaystyle \frac{1}{16}$ & $\displaystyle \frac{1}{32}$&$\displaystyle \frac{1}{16}$    & $\displaystyle \frac{3}{64}$ & $\displaystyle \frac{1}{64}$&$\displaystyle \frac{1}{128}$  & $\displaystyle \frac{1}{32}$ & $\displaystyle \frac{3}{128}$  & $\displaystyle \frac{7}{512}$&$\displaystyle \frac{1}{256}$  & $\displaystyle \frac{1}{512}$        \cr
\hline
\end{tabular}
\caption{The maximal coherence intensities $I_n^{max}$ of the $n$-order coherence and the rank $N_n$ 
of $\tilde \rho^{(n)}$ for the different numbers of nodes $N$ in a spin system. }\label{Table1}
\end{table}
Regarding the  minimum of any non-zero-order coherence intensity, its value is obvious:
\begin{eqnarray}
I_n^{min} = 0.
\end{eqnarray}

\section{Evolution of  coherences}
\label{Section:ev}

\subsection{Conservation  laws}

First of all we remind a famous conservation law which holds for any evolutionary quantum system. 

{\bf Proposition 3.}
The sum of all coherence intensities conserves:
\begin{eqnarray}\label{Lrho2}
\frac{d}{d t} \sum_{n=-N}^N I_n = \frac{d}{d t}{\mbox{Tr}} \Big( \rho^{(n)}\rho^{(-n)}\Big) =0.
\end{eqnarray}

{\it Proof.} In fact,
{ consider the Liouvile equation 
\begin{eqnarray}\label{L}
i \frac{d \rho}{dt} =[\rho,H].
\end{eqnarray}
Using this equation we have 
\begin{eqnarray}
i{\mbox{Tr}}\frac{d \rho^2}{dt} = {\mbox{Tr}} [\rho^2,H] =0.
\end{eqnarray}
Therefore
\begin{eqnarray}
{\mbox{Tr}}\rho^2 = {\mbox{Tr}}\left(\sum_{n=-N}^N \rho^{(n)}\rho^{(-n)}\right) = \sum_{n=-N}^N  {\mbox{Tr}} (\rho^{(n)}\rho^{(-n)}) = \sum_{n=-N}^N I_n\equiv const.
\end{eqnarray}
which is equivalent to 
 eq.(\ref{Lrho2})}. $\Box$
 
 In addition, if the system evolves under the Hamiltonian commuting with $I_z$,
 \begin{eqnarray}\label{comm}
 [H,I_z]=0,
 \end{eqnarray}
 then there is a family  of  conservation laws specified as follows.
 
{\bf Consequence.}
If (\ref{comm}) holds then all coherences conserve, i.e.
\begin{eqnarray}\label{cohI}
\frac{dI_n}{dt} = 0,\;\;\; |n|\le N .
\end{eqnarray}

{\it Proof.}

From eq.(\ref{L}) we have
\begin{eqnarray}
i \rho^{(n)} \frac{d \rho}{dt} + i \frac{d \rho}{dt}\rho^{(-n)}  = \rho^{(n)} [ H,\rho]  +
 [H,\rho] \rho^{(-n)} .   
\end{eqnarray}
The trace of this equation reads
\begin{eqnarray}\label{Tr0}
&& {\mbox{Tr}} \left(i \rho^{(n)} \frac{d \rho}{dt} + i \frac{d \rho}{dt}\rho^{(-n)} \right) =
i \frac{d}{dt } {\mbox{Tr}} \Big( \rho^{(n)} \rho^{(-n)}\Big) \equiv \\\nonumber
&&
i \frac{d I_n}{dt }   =  {\mbox{Tr}}\Big(\rho^{(n)}  H\rho-\rho  H\rho^{(n)}\Big)  -
{\mbox{Tr}}\Big( \rho H \rho^{(-n)}-\rho^{(-n)} H \rho\Big).
\end{eqnarray}
We can introduce factors $e^{i \phi I_z}$ and $e^{-i \phi I_z} $ under the trace, substitute expansion (\ref{RhoC}) for $\rho$  and use   commutation relation (\ref{comm}). Then we have
\begin{eqnarray}\label{TrTr}
 &&
 {\mbox{Tr}} \Big(e^{i \phi I_z} (\rho^{(n)}  H\rho-\rho  H\rho^{(n)} )e^{-i \phi I_z}\Big)  
 -{\mbox{Tr}} \Big(e^{i \phi I_z} (\rho H \rho^{(-n)}-\rho^{(-n)} H \rho)e^{-i \phi I_z}\Big) =\\\nonumber
 &&
 \sum_{k=-N}^N \left({\mbox{Tr}} \Big( e^{i \phi (n+k) } (\rho^{(n)}  H\rho^{(k)} -\rho^{(k)}  H\rho^{(n)})\Big)   -
 {\mbox{Tr}}\Big( e^{i \phi (k-n)} (\rho^{(k)} H \rho^{(-n)}-\rho^{(-n)} H \rho^{(k)})\Big) \right).
\end{eqnarray}
Since this trace must be independent on $\phi$ we have $k=-n$ and $k=n$ in the first and the second trace respectively. 
Therefore expression (\ref{TrTr}) is identical to zero and eq.(\ref{Tr0}) yields
set of conservation lows  (\ref{cohI}).
$\Box$

Equalities (\ref{cohI}) represent the set of conservation laws associated with the dynamics of a spin system under the Hamiltonian $H$ 
commuting with $I_z$.

\subsection{On map  $\rho^{(n)}(0) \to \rho^{(n)}(t)$ }

Here we derive an important consequence of conservation laws (\ref{cohI}) describing the dependence of the elements of the evolutionary matrix $\rho^{(n)}(t)$ on the elements of the initial matrix $\rho^{(n)}(0)$.
First of all we notice that the Hamiltonian commuting with $I_z$ has the following block structure:
\begin{eqnarray}\label{Hn}
H=\sum_{l=0}^N H^{(l)},
\end{eqnarray}
where the block $H_l$ governs the dynamics of states with $l$ excited spins ($l$-excitation block). 
Then any matrix $\rho^{(n)}$ can be also represented as 
\begin{eqnarray}
\rho^{(n)}=\sum_{l=0}^{N-n} \rho^{(l,l+n)},\;\;
\rho^{(-n)}=\sum_{l=n}^{N} \rho^{(l,l-n)},\;\;n=0,1,\dots,N.
\end{eqnarray}
Then, introducing the evolution operators 
\begin{eqnarray}
V(t)=e^{-i H t},\;\;\; V^{(l)}(t)=e^{-i H^{(l)} t},
\end{eqnarray}
we can write the evolution of the density matrix as 
\begin{eqnarray}
&&
\rho(t)=V(t) \rho(0) V^+(t) = \sum_{n=-N}^N V(t) \rho^{(n)}(0) V^+(t) =\\\nonumber
&&
\sum_{n=0}^N \sum_{l=0}^{N-n}  V^{(l)}(t) \rho^{(l,l+n)}(0) (V^{(l+n)}(t))^+ +
\sum_{n=-N}^{-1} \sum_{l=n}^{N}  V^{(l)}(t) \rho^{(l,l-n)}(0) (V^{(l-n)}(t))^+ .
\end{eqnarray}
Since the operators $V^{(l)}$ do not change the excitation number, we can write 
\begin{eqnarray}\label{In0}
&&
\rho(t) =\sum_{n=-N}^N \rho^{(n)}(t),\\\label{In}
&&
\rho^{(n)}(t) = \sum_{l=0}^{N-n}  V^{(l)}(t) \rho^{(l,l+n)}(0) (V^{(l+n)}(t))^+\equiv P^{(n)} \left[t, \rho^{(n)}(0)\right],\\\nonumber
&&
\rho^{(-n)} = (\rho^{(n)}(t))^+ = \sum_{l=n}^{N}  V^{(l)}(t) \rho^{(l,l-n)}(0) (V^{(l-n)}(t))^+\equiv P^{(-n)} \left[t, \rho^{(-n)}(0)\right],
\end{eqnarray}
where we introduce the linear evolutionary  operators $P^{(n)}$ ($P^{(-n)}$) mapping 
 the matrix $\rho^{(n)}(0)$  ($\rho^{(-n)}(0)$)  into the evolutionary  matrix  $\rho^{(n)}(t)$ ($\rho^{(-n)}(t)$) responsible for the same $n$-order ($(-n)$-order) coherence, i.e., 
 the operator $P^{(n)}$ applied to the matrix of the $n$-order coherence doesn't generate coherences of different order. We notice that, in certain sense,  formulas (\ref{In}) are similar to the Liouville representation \cite{Fano}. Hereafter we do not write $t$ in the 
arguments of $P^{(n)}$ for simplicity.

\section{Coherence transfer from  sender to receiver}
\label{Section:cohtr}

\subsection{Coherence transfer as map $\rho^{(S)}(0)\to \rho^{(R)}(t)$}
\label{Section:map}
Now we consider the process of the coherence transfer from the M-qubit sender ($S$) to the M-qubit receiver ($R$) connected by the transmission line ($TL$).
The receiver's density matrix reads
\begin{eqnarray}\label{rhoR}
\rho^R(t)={\mbox{Tr}}_{/R}\rho(t)= \sum_{n=-M}^M \rho^{(R;n)}(t),
\end{eqnarray}
where the trace is taken over all the nodes of the quantum system except the receiver, and $\rho^{(R;n)}$ means the submatrix of $\rho^{(R)}$ contributing into the $n$-order coherence.

To proceed further, we consider the tensor product initial state 
\begin{eqnarray}
\rho(0)=\rho^{(S)}(0)\otimes \rho^{(TL,R)}(0),
\end{eqnarray}
Obviously
\begin{eqnarray}
\rho^{(n)}(0) = \sum_{n_1+n_2=n} \rho^{(S;n_1)}(0)\otimes \rho^{(TL,R;n_2)}(0),
\end{eqnarray}
where  $\rho^{(S;n)}$ and  $\rho^{(TL,R;n)}$ are matrices contributing to the $n$-order coherence of, respectively, $\rho^{(S)}$ and $\rho^{(TL)}$.
Using expansion (\ref{In0})  and operators $P^{(n)}$ defined in (\ref{In}) we can write
\begin{eqnarray}
\rho^{(R)} = {\mbox{Tr}}_{/R} \sum_{n=-N}^N P^{(n)} \left[\rho^{(n)}(0)\right]= 
{\mbox{Tr}}_{/R} \sum_{n=-N}^N \sum_{n_1+n_2=n} P^{(n)} \left[\rho^{(S;n_1)}(0)\otimes \rho^{(TL,R;n_2)}(0)\right].
\end{eqnarray}
Next we need the following Proposition.

{\bf Proposition 4.}
The partial trace of matrix $\rho$ does not mix coherences of different order and, in addition, 
\begin{eqnarray}\label{PT}
{\mbox{Tr}}_{/R} \rho^{(n)} = 0,\;\; |n|>M,
\end{eqnarray}

{\it Proof.}
We split the whole multiplicative basis of quantum state into the $2^M$-dimensional sub-basis $B^{(R)}$ of the receiver's states and the $2^{N-M}$-dimensional  sub-basis of  the subsystem consisting of the  sender and the  transmission line $B^{(S,TL)}$,
i.e., $|i\rangle = |i^{S,TL}\rangle \otimes |i^R\rangle $. Then
elements of the density matrix $\rho$ are enumerated by the double indexes $i=(i^{S,TL},i^R)$ and $j=(j^{S,TL},j^R)$, i.e.,
\begin{eqnarray}
\rho_{ij}=\rho_{(i^{S,TL},i^R),(j^{S,TL},j^R)}. 
\end{eqnarray}
Then eq.(\ref{rhoR}) written in components reads
\begin{eqnarray}
\rho^{(R)}_{i^Rj^R} =  {\mbox{Tr}}_{/R} \rho = \sum_{i^{S,TL}} \rho_{(i^{S,TL},i^R),(i^{S,TL},j^R)}.
\end{eqnarray}
Therefore the coherences in the matrix $\rho^{(R)}$ are formed only by the transitions in the subspace spanned by $B^{(R)}$.  Therefore, the matrix $\rho^{(R;n)}$ forming the $n$-order coherence of the receiver consists of the elements included into the $n$-order coherence of the whole quantum system. Consequently, trace does not mix coherences. 

Since the receiver is an $M$-qubit subsystem, it can form only the coherences of order $n$ such that  $|n|\le M$, which agrees with justifies condition (\ref{PT}). 
$\Box$

This Proposition allows us to conclude that
\begin{eqnarray}\label{Rn}
\rho^{(R;n)} =  {\mbox{Tr}}_{/R} \sum_{n_1+n_2=n} P^{(n)}\left[ \rho^{(S;n_1)}(0)\otimes \rho^{(TL,R;n_2)}(0)\right],\;\; |n|\le M.
\end{eqnarray}

Formula (\ref{Rn}) shows that, in general,  all the coherences of $\rho^{(S;n)}$ are mixed in any particular order coherence of the  receiver's density matrix $\rho^R$. However,
this is not the case if the initial state $\rho^{TL,R}(0)$  consists of elements contributing only to the zero-order coherence. 
Then (\ref{Rn}) gets the form
\begin{eqnarray}\label{Rn2}
\rho^{(R;n)} =  {\mbox{Tr}}_{/R} \Big( P^{(n)} \Big[\rho^{(S;n)}(0)\otimes \rho^{(TL,R;0)}(0)\Big]\Big),\;\; |n|\le M.
\end{eqnarray}
In this case the elements contributing to the $n$-order coherence of $\rho^S(0)$ contribute  only to the $n$-order coherence of $\rho^R(t)$. 

\subsection{Restoring of  sender's state at  receiver's side}
\label{Section:selecting}
In Sec.\ref{Section:map} we show that, although the coherences of the sender's initial state are properly separated in the
receiver's state, the elements contributing to the particular $n$-order coherence of $\rho^S_0$ are mixed in $\rho^R_n$. But we would like to separate  the elements of $\rho^S_0$ in $\rho^R(t)$, so that, in the ideal case,:
\begin{eqnarray}\label{rhoij}
&&\rho^R_{ij}(t) = f_{ij}(t) \rho^S_{ij},\;\;(i,j)\neq (2^M,2^M),\\\nonumber
&&\rho^R_{2^M2^M}(t) = 1- \sum_{i=1}^{2^M-1} f_{ii}(t) \rho^S_{ii}.
\end{eqnarray}
We refer to the state with elements  satisfying (\ref{rhoij}) as a completely restored state.
Perhaps, relation (\ref{rhoij}) can not be realized for all elements of $\rho^R$, in other words, the complete sender's state restoring  is  impossible, in general case. 
However, the simple case of a complete restoring is the transfer of the one-qubit sender state to the one-qubit receiver because in this case there is only one  element $\rho^S_{12}$  in $\rho^S$ contributing to the first order coherence in $\rho^R$ and one independent element $\rho^S_{11}$ contributing to the zero-order coherence.
In addition, we can notice that  the highest order coherences have the form (\ref{rhoij}) in general case, because there is only one element of the density matrix contributing to the $\pm M$-order   coherence.
Regarding the other coherences, 
we can try to partially restore  at least  some of the elements using the local unitary transformation  at the receiver side.

\subsubsection{Unitary transformation of extended receiver as state-restoring tool} 
\label{Section:U}

Thus we can use the unitary transformation at the receiver to (partially) restore the initial sender's state $\rho^{(S)}(0)$ in the density matrix $\rho^{(R)}(t)$ at some time instant $t$ in the sense of definition (\ref{rhoij}).
It is  simple to estimate that the number of parameters in the unitary transformation $U^{(R)}$ of the receiver itself   is not enough to restore all the elements of the density matrix $\rho^{(S)}(0)$. To make the complete restoring possible we must increase the number of parameters in the unitary transformation  by extending the receiver to $M^{(ext)}>M$ nodes and use the transformation $U^{(ext)}$ of this extended receiver to restore the state $\rho^{(S)}(0)$.

Thus we consider the $M^{(ext)}$-dimensional extended receiver
and  require that the above mentioned unitary  transformation does not mix different submatrices $\rho^{(n)}$. This is possible if $U$ commutes with the  $z$-projection of the total extended receiver's spin momentum.   
In this case the matrix $\rho^R$ can be obtained from $\rho$ in three steps: (i) reducing $\rho(t)$ to the density matrix of the extended receiver 
$\rho^{R_{ext}}(t)$, (ii) applying the restoring unitary transformation $U^{(ext)}$ and (iii) reducing the resulting density matrix $U^{(ext)}\rho^{R_{ext}}(t)(U^{(ext)})^+$ to 
$\rho^{R}$. 
To find out the general form of the unitary transformation we consider this transformation in the basis constructed on the matrices $I^{\pm}_j$ and $I_{zj}$.
This basis reads:

for the one-qubit subsystem ($i$th qubit of the whole quantum system),
\begin{eqnarray}\label{B1}
B^{(i)}: E, I_{zi}, I^+_i, I^-_i;
\end{eqnarray}
for the two-qubit subsystem (the $i$th and $j$th qubits),
\begin{eqnarray}\label{B2}
B^{(ij)}=B^{(i)}\otimes B^{(j)};
\end{eqnarray}
for the three-qubit subsystem (the $i$th, $j$th and $k$th qubits),
\begin{eqnarray}\label{B3}
B^{(ijk)}=B^{(ij)}\otimes B^{(k)};
\end{eqnarray}
for the four-qubit subsystem (the $i$th, $j$th, $k$th and $m$th  qubits),
\begin{eqnarray}\label{B4}
B^{(ijkm)}=B^{(ij)}\otimes B^{(km)},
\end{eqnarray}
and so on.
The elements of  the basis commuting with $I_z$ are formed by the pairs   $I^+_p I^-_q$ and by the diagonal matrices $I_{zk}$, $E$.
Thus,  the one-qubit basis (\ref{B1})  involves two elements commuting with $I_z$: 
\begin{eqnarray}\label{B1U}
B^{(C;i)}: E, I_{zi}.
\end{eqnarray}
The two-qubit basis (\ref{B2})  involves  $6$ such elements:
\begin{eqnarray}\label{B2U}
B^{(C;ij)}: E, \;\;I_{zi},\;\; I_{zj}, \;\;I_{zi} I_{zj}, \;\;I^+_i I^-_j,\;\; I^+_j I^-_i.
\end{eqnarray}
The three-qubit basis (\ref{B3}) involves  20 such elements:
\begin{eqnarray}\label{B3U}
B^{(C;ijk)}: E,\;\; I_{zp},\;\; I_{zp} I_{zs},\;\; I_{zi} I_{zj}I_{zk},\;\; I^+_p I^-_s,I^+_p I^-_s I_{zr}, \;\; p,s,r\in \{i,j,k\}, \;r\neq p \neq s .
\end{eqnarray}
The four-qubit basis (\ref{B4}) involves 70 such elements:
\begin{eqnarray}\label{B4U}
B^{(C;ijkm)} &:& E, \;\;I_{zp}, \;\;  I_{zp} I_{zs},\;\;  I_{zp} I_{zs}I_{zr},\;\;I_{zi} I_{zj} I_{zk} I_{zm},\;\; 
I^+_p I^-_s,\;\;I^+_p I^-_s I_{zr},\;\;I^+_p I^-_s I_{zr} I_{zq}, \\\nonumber
&&
I^+_p I^-_s I^+_r I^-_q,\;\;p,s,r,q \in \{i,j,k,m\},\;\; p\neq s \neq r \neq q  ,
\end{eqnarray}
and so on.
However, there is a common phase which can not effect the elements of the density matrix. Therefore, the number of parameters in the above unitary transformations which can effect the density-matrix elements is less then the dimensionality of the bases (\ref{B1U}-\ref{B4U}) by one.

\section{Particular model}
\label{Section:model}
As a particular model, 
we consider the spin-1/2 chain with two-qubit sender and receiver and the   tensor product initial state
\begin{eqnarray}\label{in2}
\rho(0)=\rho^S(0) \otimes \rho^{TL,R}(0),
\end{eqnarray}
where $\rho^S(0)$ is an arbitrary  initial state of the sender 
 and $\rho^{TL,R}(0)$ is the initial  thermal equilibrium state of the transmission line and receiver,
 \begin{eqnarray}
 \label{inTLB}
\rho^{TL,B} &=&\frac{e^{bI_{z}}}{Z},\;\;Z=\left(2 \cosh\frac{b}{2}\right)^{N-2},
 \end{eqnarray}
 where $b=\frac{1}{k T}$, $T$ is temperature and $k$ is the Boltzmann constant. Thus, both $\rho^{(S)}$ and 
$\rho^{(R)}$ are $4\times 4$ matrices.

Let the evolution of  the spin chain be governed by   the nearest-neighbor $XX$-Hamiltonian \cite{Mattis}
\begin{eqnarray}\label{XY}
H=\sum_{i=1}^{N-1} D (I_{ix}I_{(i+1)x} +I_{iy}I_{(i+1)y}), 
\end{eqnarray}
where $D$ is a coupling constant. Obviously, $[H,I_z]=0$. 
Using  the Jordan-Wigner transformations \cite{JW,CG} we can derive the  explicit formula for the density matrix of the two-qubit receiver (\ref{rhoR}) 
but we do not represent the details of this derivation for the sake of brevity.

To proceed further, let us write formulas (\ref{Rn}) contributing into each particular coherence as follows. 
 For the zero order coherence we have 
\begin{eqnarray}\label{coh0}
\rho^{(R;0)}_{ij}&=& \alpha_{ij;11} \rho^S_{11} + \alpha_{ij;22} \rho^S_{22} +
\alpha_{ij;33} \rho^S_{33} + \alpha_{ij;44} \rho^S_{44} + 
\alpha_{ij;23} \rho^S_{23} + \alpha_{ij;32} (\rho^S_{23})^* 
,\\\nonumber
&&(i,j)= (1,1),(2,2),(3,3),(2,3)\\\nonumber
\rho^{(R;0)}_{44} &=& 1- \sum_{i=1}^3 \rho^R_{ii},\;\;\alpha_{ii;32}=\alpha_{ii;23}^*,
\end{eqnarray}
there are $12$ real parameters $\alpha_{ii;jj}$, $i=1,2,3$, $j=1,2,3,4$, and $9$ complex parameters $\alpha_{ii;23}$, $i=1,2,3$, $\alpha_{23;ii}$, $i=1,2,3,4$, $\alpha_{23;23}$ and $\alpha_{23;32}$, i.e., 30 real parameters.
For the first order coherence:
\begin{eqnarray}\label{coh1}
(\rho^R_1)_{ij}= \alpha_{ij;12} \rho^S_{12} + \alpha_{ij;13} \rho^S_{13} +
\alpha_{ij;24} \rho^S_{24} + \alpha_{ij;34} \rho^S_{34},\;\;
(i,j)= (1,2),(1,3),(2,4),(3,4),
\end{eqnarray}
there are 16 complex parameters, or 32 real ones.
Finally, for the second order coherence we have
\begin{eqnarray}\label{coh2}
\rho^R_{14}= \alpha_{14;12} \rho^S_{14},
\end{eqnarray}
there is one complex parameter (two real ones).
In all these formulas, 
 $\alpha_{ij;nm}$ are defined by the interaction Hamiltonian and they depend on the time $t$. 

 \subsection{Simple example of $\rho^{(S;1)}$-restoring}
 
 We see that there are 64 real parameter we would like to adjust in eqs.(\ref{coh0}-\ref{coh2}). 
For the purpose of complete  restoring   of an arbitrary state we need the extended receiver of  $M=4$ nodes so that  the number of the effective parameters in the unitary transformation described in Sec.\ref{Section:U} would be 69. 
However, for the sake of simplicity, here we use the unitary transformation of the two-qubit receiver to perform a complete  restoring of the  $\pm1$-order coherence  matrices $\rho^{(S;\pm 1)}(0)$ of a special form, namely 
\begin{eqnarray}\label{inS}
\rho^{(S;1)} + \rho^{(S;-1)} = 
\left(
\begin{array}{cccc}
0&a&a&0\cr
a^*&0&0&a\cr
a^*&0&0&0\cr
0&a^*&0&0
\end{array}
\right).
\end{eqnarray}
The unitary transformation  constructed on the basis (\ref{B2U}) reads:
\begin{eqnarray}\label{U2q}
U=e^{i \phi_1 ( I_1^+I_2^- + I_1^-I_2^+)} e^{ \phi_2 ( I_1^+I_2^- - I_1^-I_2^+)} e^{i \Phi},
\end{eqnarray}
where $\Phi={\mbox{diag}}(\phi_3,\dots,\phi_6)$ is a diagonal matrix and $\phi_i$, $i=1,\dots,6$, are arbitrary real parameters. 
Eqs. (\ref{coh1}) reduce to 
\begin{eqnarray}\label{coh1ex}
(\rho^R_1)_{ij}=\alpha_{ij} a ,\;\; \alpha_{ij}=    \alpha_{ij;12} +  \alpha_{ij;13} +
\alpha_{ij;24},\;\;
(i,j)= (1,2),(1,3),(2,4),(3,4).
\end{eqnarray}
We consider the chain of  $N=20$ nodes and  set $b=10$. The time instant for the state registration at the receiver is chosen by the requirement to maximize the maximal-order  coherence intensity (the second order in this model)  because this intensity has the least maximal possible value according to (\ref{order}). This time instance was found numerically and it equals  $D t=24.407$. 

Next,
using the parameters $\phi_i$ of the unitary transformation (\ref{U2q}) we can put zero the coefficient
$\alpha_{34}$ and thus obtain the completely restored matrices $\rho^{(R;\pm1)}$  in the form
\begin{eqnarray}\label{Rt}
\rho^{(R;1)} + \rho^{(R;-1)} = 
\left(
\begin{array}{cccc}
0&\alpha_{12} a&\alpha_{13}a&0\cr
\alpha_{12}^*a^*&0&0&\alpha_{24}a\cr
\alpha_{13}^*a^*&0&0&0\cr
0&\alpha_{24}^*a^*&0&0
\end{array}
\right).
\end{eqnarray}
The appropriate values of the parameters $\phi_i$ are following:
\begin{eqnarray}
\phi_1=2.41811,\;\;\phi_2=1.57113,\;\;\phi_k=0,\;\;k=2,\dots,6.
\end{eqnarray}
Therewith, 
\begin{eqnarray}
\alpha_{12}=0.00021 + 0.63897 i,\;\;\;\alpha_{13}=0.00010 - 0.30585 i,\;\;\alpha_{24}=0.00010-0.30582 i .
\end{eqnarray}
Thus, using the unitary transformation of the receiver we restore the sender's initial  matrices $\rho^{(S;\pm1)}(0)$ in the sense of definition (\ref{rhoij}).
This result holds for arbitrary admittable initial matrices $\rho^{(S;0)}(0)$ and $\rho^{(S;2)}(0)$.

\section{Conclusion}
\label{Section:conclusion}

The MQ  coherence intensities are the characteristics of a density matrix which can be measured in MQ NMR experiments. We show that the coherences evolve independently if only 
the Hamiltonian governing the spin dynamics conserves the total $z$-projection of the  spin momentum. This is an important property of quantum coherences which allows us to store  them in the sense that the family of the density-matrix elements contributing into a particular-order coherence do not intertwist with other elements during evolution.  In addition, if we connect the spin system with formed coherences (called sender in this case) to the transmission line and receiver we can transfer these coherences without mixing them if only the initial state of $\rho^{(TL,R)}(0)$ has only the zero-order coherence.  

We also describe the restoring method which could allow (at least partially) to reconstruct the sender's initial state. This state-restoring is based on the unitary transformation at the receiver side involving, in general,  the so-called extended receiver with the purpose to enlarge the number of parameters in the unitary transformation. The partial state-restoring  of  two-qubit receiver via the unitary transformation on it is performed as a simplest example. Examples of  more accurate restoring  involving the extended receiver require
large technical work and will be done in a specialized paper. 

This work is partially supported by the Program of RAS ''Element base of quantum
computers'' and by the Russian Foundation for Basic Research, grants No.15-07-07928 and 
16-03-00056.

\end{document}